\documentclass[lettersize,journal]{IEEEtran}
\usepackage{amsmath,amsfonts}
\usepackage{amsmath,stackengine}
\usepackage{algorithmic}
\usepackage{algorithm}
\usepackage{array}
\usepackage[caption=false,font=normalsize,labelfont=sf,textfont=sf]{subfig}
\usepackage{textcomp}
\usepackage{stfloats}
\usepackage{url}
\usepackage{verbatim}
\usepackage{graphicx}
\usepackage{cite}
\usepackage{comment}
\usepackage{tabularx}
\usepackage{booktabs}
\newcommand{\ignore}[1]{}

\usepackage[table]{xcolor}
\usepackage[dvipsnames]{xcolor}
\usepackage{color,soul}
\definecolor{lightyellow}{RGB}{255, 242, 204}
\definecolor{lightgreen}{RGB}{217, 234, 211}
\definecolor{lightblue}{RGB}{173, 216, 230}
\definecolor{lightred}{RGB}{244, 204, 204}

\begin{document}


\title{Advanced Holographic Multi-Antenna Solutions for Global Non-Terrestrial Network Integration in IMT-2030 Systems}

\author{Alfredo Nunez-Unda, Angelo Vera-Rivera~\IEEEmembership{Member,~IEEE,}, Nuwan Balasuriya, and Ekram Hossain,~\IEEEmembership{Fellow,~IEEE}
\thanks{A. Nunez-Unda is with the Department of Electrical and Computer Engineering, University of Manitoba, Winnipeg, Manitoba, Canada (email: nuneza@myumanitoba.ca), and with the Faculty of Electrical and Computer Engineering, GICOM, Escuela Superior Politécnica del Litoral, ESPOL, Guayaquil, Ecuador. A. Vera-Rivera, N. Balasuriya, and E. Hossain are with the Department of Electrical and Computer Engineering, University of Manitoba, Winnipeg, Manitoba, Canada (e-mail: umbalasd@myumanitoba.ca, \{angelo.verarivera, ekram.hossain\}@umanitoba.ca). Corresponding author: Ekram Hossain.}}

\maketitle


\begin{abstract}

Sixth-generation (6G) networks are expected to provide ubiquitous connectivity across terrestrial and non-terrestrial domains. This will be possible by integrating non-terrestrial networks (NTNs) to extend coverage to underserved areas. Antennas are central to this vision, with multiple-input multiple-output (MIMO) technologies receiving the most attention due to their ability to exploit spatial multiplexing to improve link capacity and reliability. However, conventional MIMO can consume significant energy, as each antenna element typically requires an independent RF chain. This limitation is particularly critical in non-terrestrial systems, where onboard energy resources are limited. Holographic MIMO (HMIMO) has emerged as a promising alternative in this context. These systems are based on theoretically continuous apertures, where radiation is generated through controlled modulation of surface impedance. This enables beamforming mechanisms with significantly fewer RF chains, reducing power consumption. In this work, we make the case for HMIMO as a suitable candidate for NTN integration within IMT-2030 systems. We discuss its advantages over conventional MIMO and present a case study of HMIMO integration in LEO-based multi-user communication.
\end{abstract}


\section*{Introduction}
\begin{figure*}[!t]
\begin{center}
\includegraphics[width=0.7\textwidth]{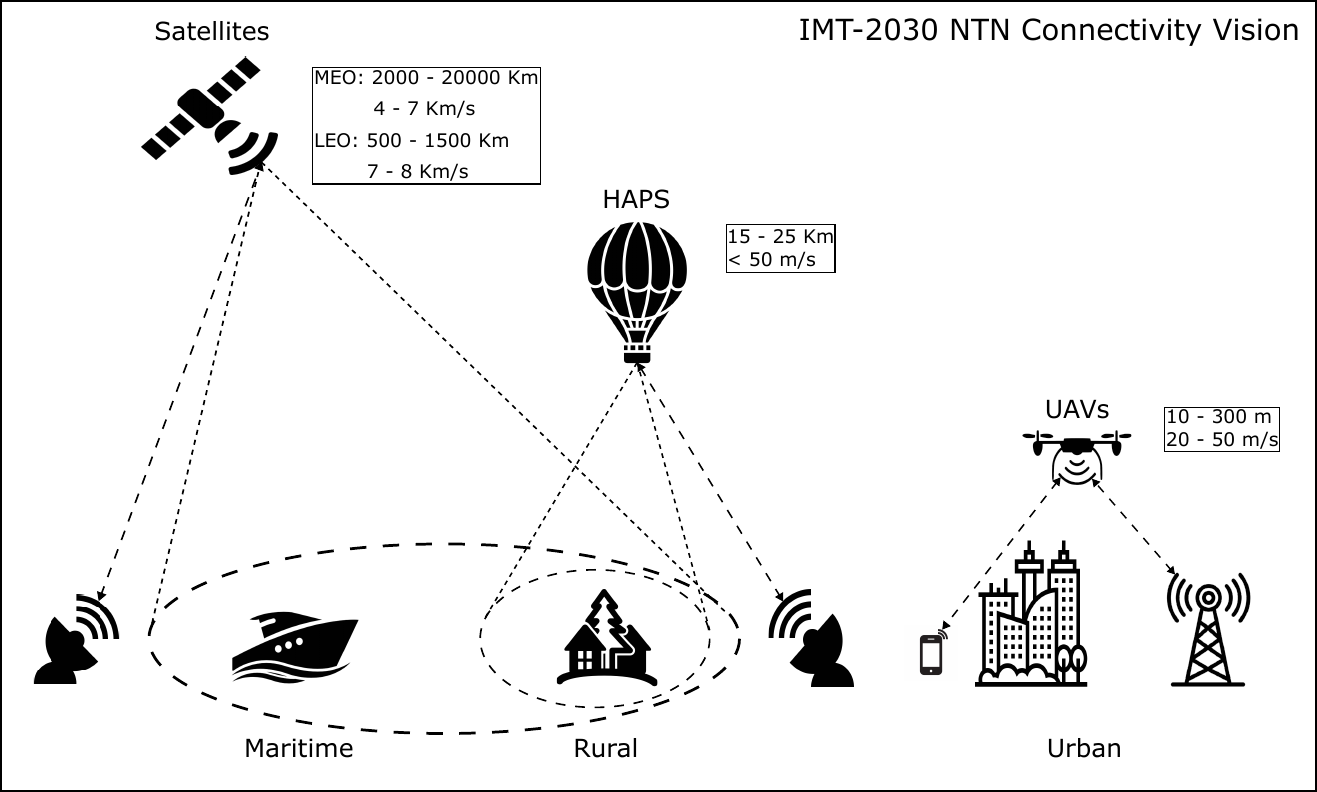}
\end{center}
\caption{Illustration of NTN integration as envisioned in IMT-2030, where satellites, HAPS, and UAVs complement terrestrial coverage infrastructure to serve maritime, rural, and urban areas.}
\label{NTN}
\end{figure*}

\IEEEPARstart{G}{lobal} coverage is a core objective of sixth-generation (6G) mobile broadband systems. As outlined in the IMT-2030 framework\footnote{IMT is the International Telecommunications Union (ITU)’s generic term for mobile broadband systems, including IMT-2000, IMT-Advanced, IMT-2020, and IMT-2030—also known as 3G, 4G, 5G, and 6G, respectively.}, future cellular networks are expected to provide ubiquitous, high-performance connectivity across terrestrial and non-terrestrial domains, including ground, air, sea, and space~\cite{ITU2023}. This vision relies on integrating terrestrial infrastructure with non-terrestrial networks (NTNs), including low Earth orbit (LEO) satellites, high-altitude platform systems (HAPS), and unmanned aerial vehicles (UAVs), to extend coverage to underserved areas. An illustration is shown in Fig.~\ref{NTN}. LEO systems, in particular, attract strong interest due to their favorable trade-off between propagation delay and coverage footprint compared to other non-terrestrial platforms. To date, several private LEO constellations are being deployed to provide broadband connectivity services\footnote{One of the most widely deployed satellite Internet services is Starlink, a LEO constellation operated by SpaceX, an aerospace company leading private satellite system development in North America.}. 

Antenna systems are central to the IMT-2030 vision, as they form the interface between communication systems and the propagation channel. Antenna design directly impacts coverage, capacity, interference management, and energy efficiency, which are critical for seamless connectivity. Antennas are a key enabler of 6G performance targets such as high user densities, data rates on the order of hundreds of gigabits per second, low latency, and high reliability. For NTN connectivity, recent studies highlight the potential of large-scale antenna arrays operating at higher frequency bands, often in combination with active and passive metasurfaces.

Multiple-Input Multiple-Output (MIMO) antenna technologies are a core component of modern wireless networks and a major focus of 6G research~\cite{11195786}. Large-scale MIMO exploits large antenna arrays to improve link reliability through spatial multiplexing and accurate beamforming. However, conventional implementations can consume significant
energy, as each antenna element typically requires its own radio frequency (RF) transmit chain, including power amplifiers and associated signal processing hardware. This power demand is particularly challenging in non-terrestrial systems, where onboard energy resources are limited. In addition, conventional MIMO requires antenna elements to be spaced approximately half a wavelength apart to avoid antenna coupling, resulting in large physical apertures that must comply with the mass, volume, and deployment constraints of non-terrestrial (e.g., satellite) platforms.

Despite the evident limitations, much of the NTN antenna literature remains focused on conventional large-scale MIMO. However, there is a growing interest in alternative technologies better aligned with the power and geometric constraints of satellite spacecrafts. Holographic antennas have emerged as a promising candidate in this context. These systems are based on theoretically continuous apertures, approximated in practice by quasi-continuous metasurface implementations, where electromagnetic radiation is generated through controlled modulation of surface impedance. This principle enables flexible beamforming and wavefront shaping with significantly fewer RF chains than conventional phased-array architectures. Unlike conventional MIMO arrays, holographic MIMO (HMIMO) systems do not require independent RF chains per antenna element or half-wavelength inter-element spacing. Instead, they employ densely spaced radiating elements illuminated by a small number of RF feeds, enabling radiation with reduced RF hardware complexity.

In this article, we examine advanced antenna solutions beyond conventional MIMO for NTN integration into IMT-2030 systems. In particular, we advocate holographic antenna arrays as promising candidates for NTN due to their favorable power consumption and compact physical footprint, which align well with NTN platform constraints. Our main contributions are:
\begin{enumerate}
    \item An overview of the key technical challenges in NTN connectivity.
    \item An assessment of conventional multi-antenna-based NTN coverage models and their limitations.
    \item An overview of holographic antenna technologies and their potential to overcome the limitations of conventional MIMO in enabling NTN solutions.
    \item A case study demonstrating the integration of HMIMO architectures into a Low Earth Orbit (LEO)-based communication system for multi-user cellular connectivity.
\end{enumerate}
%
%
The rest of the article is organized as follows.  The NTN challenges are outlined next which is followed by a review of 
conventional MIMO-based NTN architectures. Then, conventional MIMO and HMIMO are compared, highlighting the advantages of HMIMO for NTN integration in 6G. Next, a case study is presented illustrating the use of HMIMO in LEO-based multi-user communication systems. To this end, open challenges and future research directions for deploying HMIMO in NTN systems are discussed before the article is concluded.

%
%
\section*{Key Technical Challenges in NTN Connectivity}
NTN platforms operate at altitudes ranging from tens of meters (e.g., Uncrewed Aerial Vehicles [UAVs]) to approximately 36,000 km (e.g., geostationary Earth orbit (GEO) satellites) above the Earth's surface. In general, these platforms provide wide-area coverage with lower infrastructure deployment requirements compared to terrestrial networks, particularly in sparsely populated regions. However, the advantage comes with several non-trivial technical challenges. For clarity, these challenges are grouped into four categories, discussed next.
\subsection*{Challenge 1: Extreme Mobility Effects}
Many NTN platforms, particularly LEO and medium Earth orbit (MEO) satellites, operate at high orbital velocities, creating extreme mobility conditions within the network\footnote{LEO satellites travel at orbital velocities of approximately 7–8 km/s at altitudes of 500–1500 km. MEO satellites travel at orbital velocities of approximately 4-7 km/s at altitudes of 2000-20000 km.}. These high-velocity nodes act as moving base stations, reducing the time users spend within a beam footprint and increasing handovers, scheduling updates, and signaling overhead~\cite{Li2023}. This behavior fundamentally differs from terrestrial networks, where mobility is primarily user-driven. In NTN systems, aerial nodes' motion could amplify mobility effects, causing rapid changes in propagation conditions. This results in large Doppler shifts\footnote{LEO systems operating at typical frequencies can experience Doppler shifts on the order of tens of kHz.}, which introduce frequency offsets, channel aging, and degraded CSI. These effects lead to beam misalignment, side-lobe leakage, and overlapping edges, causing significant variability of signal-to-interference-and-noise ratio (SINR). While these impairments can be mitigated through digital or hybrid beamforming techniques with frequent updates, this comes at the cost of increased precoding complexity, as well as onboard processing and energy requirements~\cite{Kim2025}.
\subsection*{Challenge 2: Long-Distance Propagation and Atmospheric Effects}
Long-distance propagation in NTN systems leads to severe path loss, increased propagation delay, and reduced signal-to-noise ratio (SNR). In addition, NTN nodes typically operate at higher carrier frequencies compared to terrestrial cellular systems, making them more susceptible to atmospheric attenuation. Then, propagation variability comes from both (1) deterministic node motion (e.g., satellite orbits) and (2) stochastic weather effects, making channel conditions difficult to predict. The predominantly line-of-sight (LoS) nature of long-distance links constrains link budgets, reduces channel richness and spectral efficiency, and limits the effectiveness of MIMO architectures~\cite{An2026}. Weather-induced attenuation further leads to uneven SINR distributions across users, complicating precoder design and power allocation in multi-user systems. Together, these factors place significant constraints on antenna design, resource allocation, and power consumption in NTN platforms. 
%
\subsection*{Challenge 3: Onboard Resource Constraints}
NTN nodes (e.g., satellites, HAPS, and UAVs) operate under strict resource constraints, including limited onboard energy (solar panels and batteries), short mass and volume budgets, constrained thermal dissipation, and reliance on radiation-hardened hardware with usually limited performance. Resource constraints bound antenna scalability, onboard processing, large-scale coordination, and AI capabilities, often requiring offloading to ground infrastructure at the cost of increased latency. Energy limitations, in particular, are critical because they restrict transmit power and computational processing capacity, reducing the number of simultaneously served users and active subsystems on the ground. Moreover, unlike terrestrial base stations, NTN nodes cannot be upgraded after deployment, further constraining long-term system evolution.
\subsection*{Challenge 4: System-Level Performance Stability}
From a system perspective, NTN topologies are inherently time-dependent network graphs. Unlike terrestrial networks, they exhibit highly dynamic structures due to moving aerial nodes, time-varying links, and intermittent ground connectivity. Consequently, routing paths for network traffic require frequent updates, leading to variable end-to-end delay, increased jitter, and reduced QoS stability. These effects could propagate to higher layers, degrading session continuity and transport-layer protocols performance~\cite{Ran2025}.
\section*{Conventional Multi-Antenna Architectures for NTN Systems}
NTN architectures account for a growing portion of the 6G research literature~\cite{11195786}. From a transmission perspective, these systems distribute electromagnetic energy across users according to the antenna aperture structure, propagation conditions, and user distribution. In LoS-dominant NTN environments, the antenna structure plays a central role in shaping the transmission geometry. In this section, we review the most common NTN architectures in the literature for multi-user mobile broadband communications.
\subsection*{Multi-Beam Architectures}
This is the primary multi-antenna architecture used in LEO broadband deployments to enable multi-user (spatial multiplexing) capabilities. It divides the satellite footprint into multiple beams, each covering a smaller ground area. While early implementations used fixed antennas per beam, modern systems employ phased arrays with analog (phase shifters), digital (precoding), or hybrid beamforming to enable multi-beam operation~\cite{Sun2024}. At its core, multi-beam implements space-division multiple access (SDMA) at the RF level, but it is not inherently multi-user and must be combined with schemes such as orthogonal frequency-division multiple access (OFDMA), rate-splitting multiple access (RSMA), or any variation of code-division multiple access (CDMA). 
\subsection*{Massive MIMO Architectures}
Massive MIMO is a widely studied approach for NTN antenna architectures due to its high array gains, which help compensate for severe path loss, and its potential to support multi-user transmission. However, increasing the number of antenna elements alone does not guarantee channel orthogonality. Favorable propagation requires either rich multi-path or sufficient user angular separation, both of which are limited in LoS-dominant NTN environments. As a result, spatial multiplexing gains are constrained, often requiring additional multiple access schemes such as OFDMA or RSMA. Moreover, scaling antenna arrays increases RF chain complexity and energy consumption, posing challenges for resource-constrained NTN platforms. To address these limitations, various approaches have been proposed, including user grouping, advanced scheduling, and machine learning-based precoding~\cite{Li2022}.
\subsection*{Distributed MIMO Architectures}
Distributed MIMO, particularly distributed massive MIMO, has been proposed to overcome the limited spatial diversity of single-node massive MIMO. In NTN platforms, clusters of interconnected aerial nodes enable the distribution of antenna arrays across multiple nodes. This spatial separation improves channel diversity, expands the coverage footprint, and reduces handover frequency~\cite{Abdelsadek2022}. In addition, distributing antennas across nodes allows power sharing, partially alleviating onboard energy constraints. Despite these benefits, achieving high array gains still requires large numbers of antenna elements, increasing RF chain complexity and power consumption. High node mobility (e.g., LEO satellites) further introduces strong Doppler effects and latency, complicating channel estimation and synchronization. Moreover, LoS-dominant conditions continue to limit achievable multiplexing gains compared to terrestrial networks.
\subsection*{RIS-Assisted Architectures}
Reconfigurable intelligent surface (RIS) technology has been proposed for next-generation mobile broadband due to its flexibility, low cost, and energy efficiency~\cite{8910627}. Its primary function is to enhance coverage by enabling artificial LoS links through programmable phase shifts. In NTN, RIS has been explored to improve link quality and mitigate interference. In LEO communications, for instance, ground-based RIS deployments have been studied for interference mitigation (e.g., from GEO and other LEO satellites) and for coverage enhancement in suburban scenarios~\cite{Zhang2025}. However, these evaluations often assume rich multi-path conditions, which are not representative of typical rural and many LoS-dominant environments. Combining distributed LEO architectures with multiple ground-based RISs could provide a more robust framework for multi-user MIMO services, while enabling improved user localization and more reliable channel estimation due to their fixed deployment.
\section*{The Role of Advanced Holographic Antenna Techniques for NTN}
\begin{table*}[htbp]
\centering
\caption{Comparison of Conventional MIMO and HMIMO for NTN–Terrestrial Integration. Green indicates aspects where HMIMO outperforms conventional MIMO, yellow indicates comparable performance, and red indicates aspects where HMIMO underperforms relative to conventional MIMO.}
\label{tab:convMIMOvsHMIMO}
\resizebox{0.92\linewidth}{!}{
\begin{tabular}{p{3.5cm} p{3.5cm} p{3.5cm} p{7cm}}
\toprule
\textbf{Aspect} & \textbf{Conventional MIMO} & \textbf{HMIMO} & \textbf{Key insights} \\
\midrule
\multicolumn{4}{l}{\textbf{\textit{RF Structure}}} \\ 
\midrule
Aperture type & Discrete array of separated radiators & Quasi-continuous aperture & \cellcolor{green!10} HMIMO enables higher antenna gains per unit area. \\
\addlinespace
Element spacing & $\lambda/2$ element spacing & sub-$\lambda/2$ element spacing & \cellcolor{green!10} Sub-wavelength spacing in HMIMO enables quasi-continuous apertures and more granular wavefront control. \\
\addlinespace
Mutual coupling & Negligible & Present & \cellcolor{red!10} In HMIMO, mutual coupling must be accounted for at sub-wavelength spacings. \\
\addlinespace
RF chain scaling & One RF chain per element with dedicated RF front-end & Few shared RF feeds with simplified RF front-end & \cellcolor{green!10} For comparable apertures, HMIMO requires fewer RF chains than conventional MIMO, reducing active RF circuitry.  \\
\addlinespace
Technology maturity & Mature & Emerging & \cellcolor{red!10} Conventional MIMO is widely deployed, while HMIMO remains in early-stage development \\
\addlinespace
\multicolumn{4}{c}{\cellcolor{green!10} \textit{Overall: HMIMO offers improved hardware scalability and reduced RF complexity for large-aperture NTN systems}} \\
\midrule
\multicolumn{4}{l}{\textbf{\textit{Propagation Behavior}}}  \\ 
\midrule
Physical footprint & Element-dependent & Element-dependent & \cellcolor{yellow!10} Similar for HMIMO and conventional MIMO with comparable apertures. \\
\addlinespace
Beamforming mechanism & Digital precoding + analog phase shifters & Digital precoding + wave-domain analog beamforming & \cellcolor{green!10} HMIMO enables hybrid beamforming without discrete phase shifters, reducing hardware complexity. \\
\addlinespace
Radiation directivity & Limited by discrete aperture and RF losses & Higher due to quasi-continuous aperture & \cellcolor{green!10} HMIMO improves aperture utilization, leading to narrower beams and lower sidelobes. \\
\addlinespace
Spatial multiplexing & Very limited under LoS-dominant conditions & Improved due to better spatial resolution & \cellcolor{green!10} HMIMO better exploits limited spatial diversity under LoS-dominant conditions.\\
\addlinespace
Doppler robustness & Well-understood Doppler compensation & Higher sensitivity under extreme mobility &  \cellcolor{red!10} HMIMO is more sensitive to Doppler shifts due to its higher directivity and narrower beams. \\ 
\addlinespace
\multicolumn{4}{c}{\cellcolor{yellow!10} \textit{Overall: HMIMO enhances wave control and directivity, but introduces sensitivity to Doppler and LoS-limited multiplexing}} \\
\midrule
\multicolumn{4}{l}{\textbf{\textit{Hardware Efficiency}}}  \\ 
\midrule
Power consumption & High (scales with number of RF chains) & Lower (reduced RF hardware) & \cellcolor{green!10} HMIMO requires fewer RF chains, improving energy efficiency.\\

\addlinespace
RF hardware losses & Higher & Lower &  \cellcolor{green!10} Reduced RF circuitry in HMIMO lowers insertion and conversion losses.\\
\addlinespace
Size and weight (SaW) & Larger & Relatively smaller & \cellcolor{green!10} 
In HMIMO, dense apertures and reduced RF hardware enable more compact and lightweight designs.\\
\addlinespace
\multicolumn{4}{c}{\cellcolor{green!10} \textit{Overall: HMIMO improves energy efficiency and reduces hardware footprint, aligning well with NTN constraints}} \\
\bottomrule
\end{tabular}
}
\end{table*}
%
In this section, we analyze the key characteristics of conventional MIMO and HMIMO. We provide a comparative assessment of both technologies from the NTN perspective and argue that HMIMO is a more suitable candidate for NTN integration with mobile broadband due to its favorable power efficiency and compact physical footprint.
\subsection*{Conventional MIMO Fundamentals}
Conventional MIMO arranges antenna elements with approximately $\lambda$/2 spacing, forming discrete radiator arrays where each element is typically connected to an individual RF chain. This leads to increased hardware complexity and power consumption as the array size grows (e.g., massive and ultra-massive MIMO). In most cases, communication is modeled under far-field conditions, where wave propagation can be approximated as planar. The system degrees of freedom are therefore limited by the number of antenna elements, i.e., $\min(N_t, N_r)$, and effective operation requires sufficiently uncorrelated channels. Beamforming is primarily implemented through digital precoding, with analog or hybrid approaches enabled by phase shifters, and relies heavily on accurate channel state information (CSI). While MIMO effectively exploits spatial diversity in multi-path environments, these conditions are difficult to satisfy in LoS-dominant and high-mobility NTN scenarios. As a result, the demand for higher data rates and device density drives the need for larger arrays, pushing conventional implementations toward practical limits.
\subsection*{HMIMO Fundamentals}
HMIMO is an advanced multi-antenna architecture where the transmitter (and/or receiver) is implemented as a densely sampled, nearly continuous electromagnetic aperture that enables wavefront control with sub-wavelength resolution~\cite{Huang2020}. The continuous aperture is theoretical, but it is approximated in practice using dense arrays with sub-wavelength spacing (e.g., $\lambda/3$--$\lambda/10$). The design is inspired by the holographic principle, where desired radiation patterns are synthesized through controlled interference between reference and object waves, with target receivers defining the intended wavefronts. Early implementations relied on static radiating surfaces, while recent designs use tunable metamaterial arrays (e.g., uniform linear arrays (ULAs) or uniform planar arrays (UPAs)) coupled through waveguides, microstrip networks, or even free-space feeds to enable dynamic beamforming via phase-profile optimization. The most prominent HMIMO architectures are discussed next \cite{Gong2024}. 
\subsubsection{Leaky-Wave-Enabled HMIMO}
A leaky-wave antenna (LWA), often referred to as a reconfigurable holographic surface (RHS), radiates energy continuously along a guiding structure through controlled leakage, enabling directive beamforming without discrete antenna elements. In LWAs, the feed injects reference waves into a densely spaced array of radiating elements with sub-wavelength spacing. These elements introduce controlled discontinuities that convert the guided waves into leaky waves, which radiate into free space. When excited, RHS shapes the resulting radiation towards desired directions.
\subsubsection{Microstrip-Enabled HMIMO}
A microstrip is a planar transmission line consisting of a conductive strip over a dielectric substrate, commonly used to guide RF signals in antenna arrays. In microstrip-enabled HMIMO, also known as dynamic metasurface antenna (DMA), multiple feeds are connected to subsets of microstrip lines, each supporting tunable patch elements with sub-wavelength spacing. Unlike RHS architectures, each feed excites only a portion of the aperture, reducing the dimensionality of the optimization problem. This structure enables more efficient analog phase control and can improve spectral efficiency.
\subsubsection{Transmissive RIS-Enabled HMIMO}
In transmissive RISs (T-RISs), also known as reconfigurable refractive surfaces (RRSs), the RF feeds are spatially separated from the metasurface. The feeds illuminate the surface with incident waves, which are then phase-modulated as they pass through the tunable elements. Each element applies a controllable phase shift to the transmitted wave, enabling wavefront shaping toward desired directions.
\subsubsection{Stacked Metasurface-Enabled HMIMO}
Stacked intelligent metasurfaces (SIMs) are a recent architecture for enabling HMIMO, consisting of multiple parallel metasurface layers. Each layer comprises tunable refractive elements, similar to RRS structures, and their stacking enables enhanced wavefront control compared to a single surface. Depending on the feed configuration, the first layer may act as an input surface (e.g., LWA or DMA), while subsequent layers operate as refractive metasurfaces. Alternatively, when feeds are externally located (near-field), all layers function as transmissive surfaces. The inter-layer spacing, ranging from sub-wavelength (reactive near-field) to several wavelengths, is a key design parameter that can be optimized for performance \cite{An2023}.
\subsection*{The Case for HMIMO over Conventional MIMO for NTN Integration}
In NTN nodes with limited onboard resources, scaling MIMO becomes prohibitive due to the large number of RF chains and digital beamformers required. In contrast, HMIMO leverages metamaterial-based apertures to synthesize radiation patterns without per-element RF chains. It enables high array gains and spatial multiplexing with sub-wavelength element spacing, reduced number of RF feeds, and a transition from fully digital to less complex hybrid beamforming~\cite{Li2025}. If a single driving factor motivates the adoption of HMIMO over conventional MIMO, it is definitely power. The primary challenge in designing multi-antenna solutions for NTN, particularly in LEO systems, is the severe path loss associated with orbital distances, which requires large antenna gains.

Achieving high gains in conventional MIMO requires a large number of radiating elements. HMIMO is more suitable in this context, as it requires significantly fewer RF chains for comparable aperture sizes. This hardware reduction enables power and computational savings that can be allocated to advanced signal processing, optimization, and channel estimation, which are also critical in NTN communication systems. Moreover, HMIMO enables simpler analog and hybrid beamforming without per-element phase shifters, offering additional degrees of freedom for beam control and side-lobe suppression. Although HMIMO alone is not sufficient to fully address the challenges of NTN communications and must be complemented by additional system-level solutions, it can, in the appropriate context, replace conventional MIMO for practical deployments. In such cases, HMIMO provides substantial savings in power, physical footprint which are key constraints in NTN systems, while shifting complexity from RF hardware to signal processing and aperture optimization.
\section*{Case Study: HMIMO-Enabled Architecture for LEO-based Multi-User Communication}
\begin{figure}[!t]
\begin{center}
\includegraphics[width=0.48\textwidth]{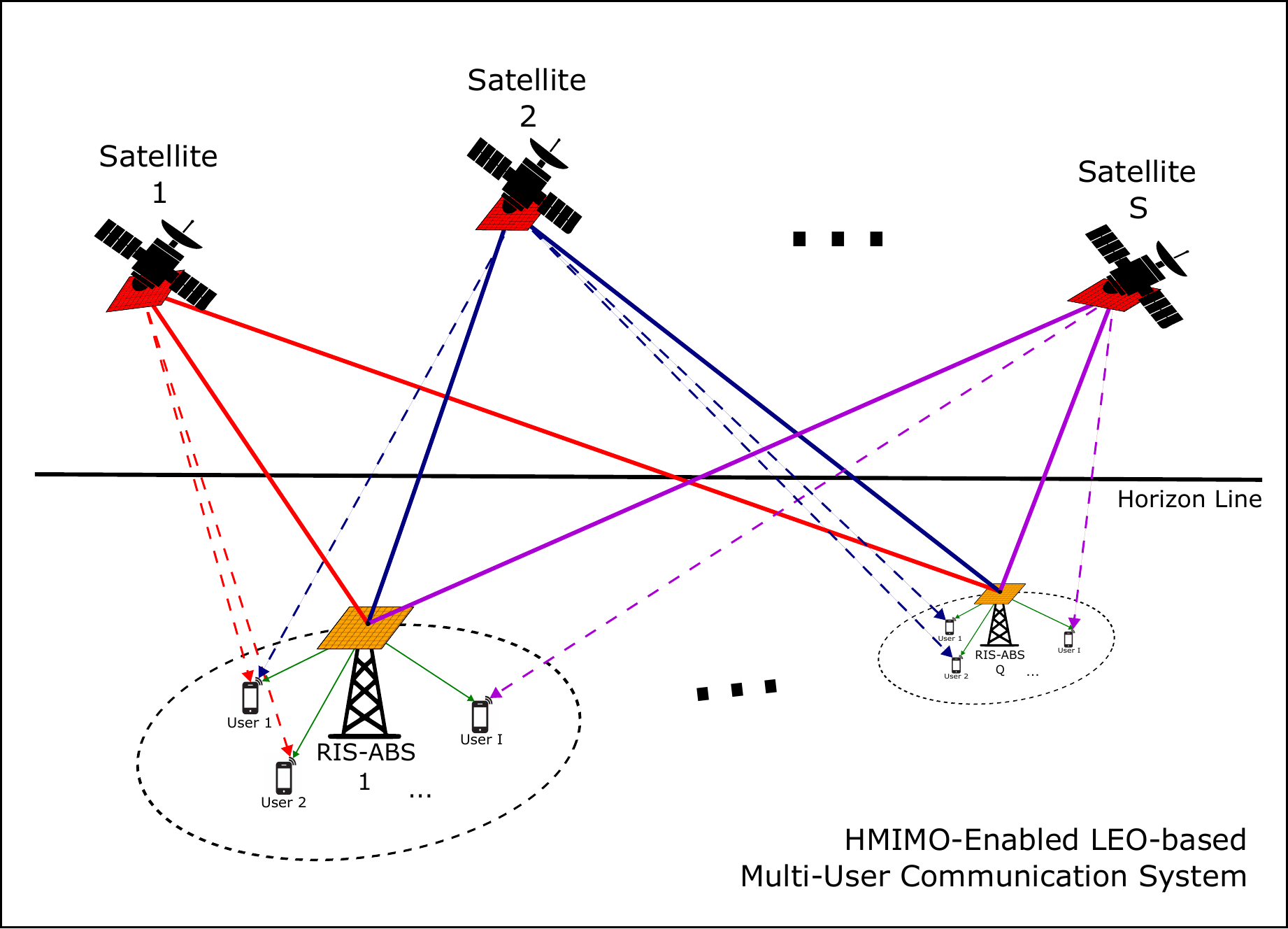}
\end{center}
\caption{Proposed HMIMO-enabled LEO satellite architecture for multi-user communication via RIS-ABS.}
\label{GenScenario}
\end{figure}

In this section, we present a case study illustrating the integration of HMIMO antennas into a LEO-based communication system for multi-user cellular connectivity. An overview of the architecture is shown in Fig.~\ref{GenScenario}.
\subsection*{System Model and Problem Formulation}
Consider a LEO constellation $\mathcal{S}=\{1,\ldots,S\}$, where each satellite is equipped with an HMIMO array comprising $N$ unimodular meta-material elements. On the ground, we assume $\mathcal{Q}=\{1,\ldots,Q\}$ base stations, each equipped with a transmissive RIS (T-RIS) with $K$ unimodular elements. We refer to this configuration as RIS-assisted base stations (RIS-ABSs). Each RIS-ABS serves a set of users $\mathcal{I}=\{1,\ldots,I\}$. The satellites are assumed to be synchronized, forming a cooperative cluster. Connectivity to the mobile core is provided via ground stations, which are omitted for clarity.

In the downlink, we assume perfect instantaneous CSI is available. We define $\mathbf{H}_{h-s}\in\mathbb{C}^{QK\times SN}$ as the channel matrix between the HMIMO arrays and the RIS-ABSs, $\mathbf{H}_{s-u}\in\mathbb{C}^{I\times QK}$ as the channel matrix between the RIS-ABSs and the users, and $\mathbf{H}_{h-u}\in\mathbb{C}^{I\times SN}$ as the direct channel between the HMIMO arrays and the users. HMIMO supports hybrid beamforming, where the digital component is represented by the precoder $\mathbf{F}$ and the analog component by the holographic beamforming matrix $\mathbf{M}$. On the ground, the matrix $\mathbf{\Upsilon}$ models the phase shifts applied by the T-RIS elements to steer the incident signals toward the users. Let $P_t$ denote the total transmit power, $\mathbf{x}$ the transmitted symbols, and $\mathbf{y}$ the received symbols. The objective is to maximize the sum-rate of all ground users via the minimum mean square error (MMSE) criterion. The resulting optimization problem, in its primitive form, is formulated as follows:
\begin{equation*}
\begin{array}{rrclcl}
\displaystyle \min_{\mathbf{M},\mathbf{F},\mathbf{\Upsilon}} & \multicolumn{3}{l}{\mathbb{E}_{\mathbf{y,x}}\left\{ \|\mathbf{y}-\mathbf{x}\|_2^2\right\}}\\
\textrm{s.t.} & \|\mathbf{Fx}\|_2^2 \leq P_t.\\
\end{array}
\end{equation*}
\subsection*{Numerical Results}
For the Monte Carlo simulation, we assume a system comprising two LEO satellites, one RIS-ABS, and two users. We assume the LEO orbit to be at an altitude of 600 km and an inter-satellite distance of $50$ km. The RIS-ABS is $20$ m tall, and the two single-antenna users are randomly distributed around it at distances of $20$-$50$ m. The transmission frequency is $f=12$ GHz, with bandwidth $B=1$ MHz. Each satellite is equipped with a single HMIMO array implemented as an RHS with $L=10$ feeds and antenna-element spacing of $\lambda/4$. The RIS-ABS is equipped with one T-RIS with antenna-element spacing $\lambda/2$. The system power is fixed at $P_t=200$ W. 

The results show the sum-rate as a function of the number of antenna elements ($N=K$) for two users under different channel conditions. Four scenarios are considered: (i) multi-path RIS-ABS–user links with negligible RHS–user links, (ii) all links LoS-dominant, (iii) multi-path propagation for both T-RIS–user and RHS–user links, and (iv) multi-path T-RIS–user links with LoS-dominant RHS–user links. As expected, the sum-rate increases with the number of elements due to higher array gain. The best performance is achieved in case (iv), followed by case (ii), as strong LoS links between the RHS and the users provide more efficient beamforming compared to scenarios with weak or multi-path-dominated RHS–user links. The resulting performance is consistent with the expected order of magnitude for satellite communication systems.
\begin{figure}[!t]
\begin{center}
\includegraphics[width= 0.55\textwidth]{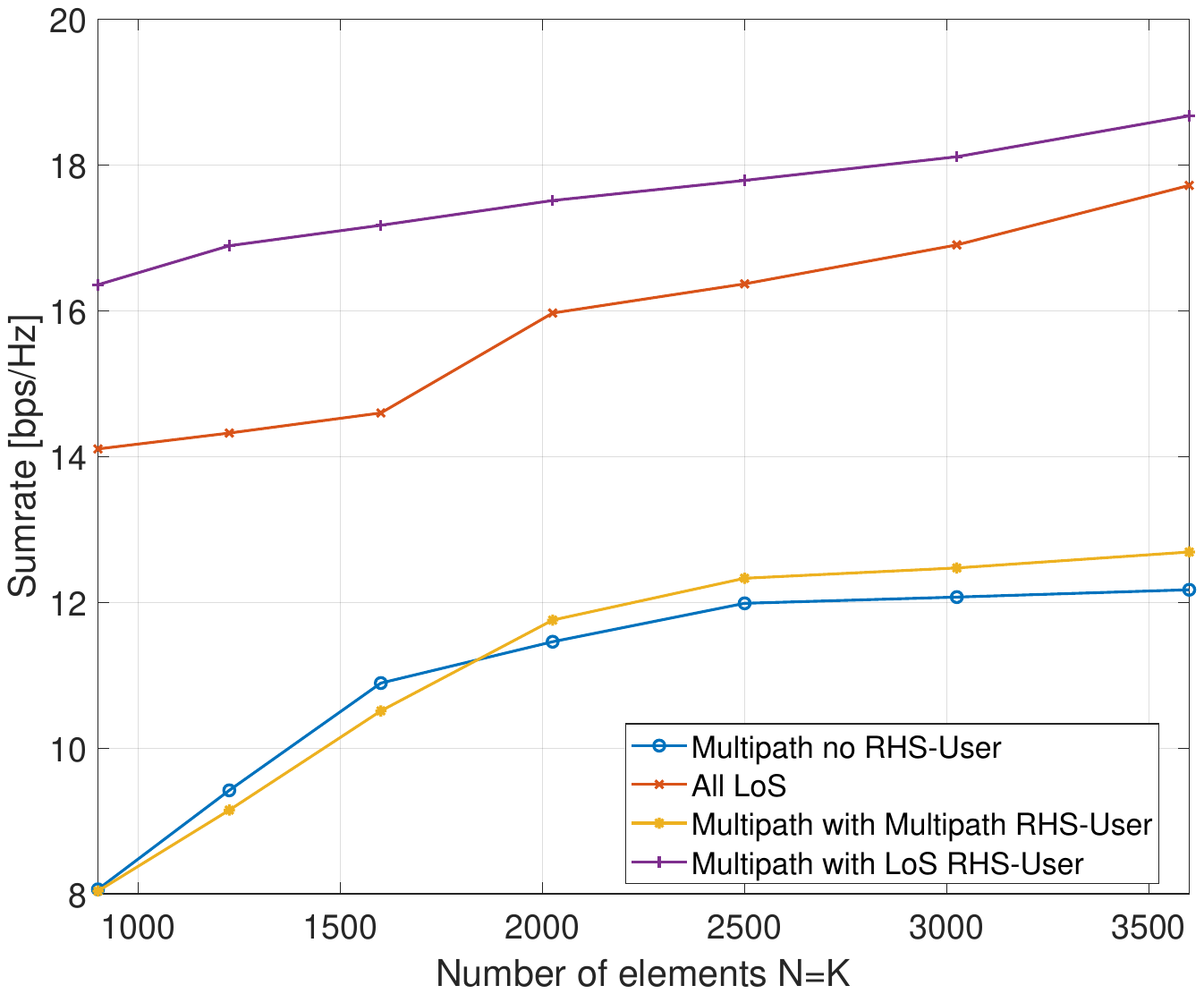}
\end{center}
\caption{Sum-rate as a function of the number of elements ($N = K$) for RHS and T-RIS under various channel conditions.}
\label{plot3}
\end{figure}
\subsection*{Implications for NTN Integration}
The value of NTN architectures, such as the one demonstrated in the case study, lies in the services and functionalities they unlock. Key opportunities for HMINO-based NTN include global-scale IoT connectivity, supporting dense, energy-efficient satellite-based deployments. Another opportunity is integrated access and backhaul (IAB), where HMIMO enables simultaneous user access and backhaul via highly directional beams. HMIMO could also support integrated sensing and communication (ISAC), facilitating global localization and tracking. In addition, it allows dynamic overlay networks for traffic offloading, coverage resilience, and load balancing across terrestrial and non-terrestrial segments. Finally, the narrow beams of HMIMO support advanced physical-layer security, including directional jamming and interference nulling. Although some of these functionalities can be supported by conventional MIMO, HMIMO enables their practical realization under the constraints of NTN systems.
\section*{Open Challenges and Future Directions}
HMIMO offers advantages for NTN connectivity but also introduces new challenges. This section highlights key issues and research directions for deploying HMIMO in NTN systems.
\subsection*{Compact and Lightweight Design}
Although HMIMO reduces RF hardware complexity and can be lighter than conventional MIMO, it still relies on large, densely packed apertures that may remain challenging for space-constrained NTN platforms. Developing more compact designs with thinner substrates and lightweight meta-materials is therefore desirable for practical NTN deployments.
\subsection*{Efficient Signal Processing}
HMIMO shifts complexity from RF hardware to signal processing optimization. In high-mobility NTN scenarios, rapidly varying channels require frequent CSI updates, which, combined with large HMIMO apertures, increase the system's orchestration complexity. Recent deep learning approaches offer promising solutions for efficient CSI estimation and signal processing in HMIMO-based solutions.
%
\subsection*{Adaptive Beamforming}
In NTN systems, severe path loss and high mobility challenge beam alignment and link reliability.  Rapid challenge variations require frequent updates, which, combined with the narrow nature of beams of HMIMO, increase sensitivity to misalignments. These factors make adaptive beamforming a critical component of HMIMO-based NTN systems, requiring responsive beam tracking and optimization to ensure reliable connectivity.
\subsection*{Multi-Band Operation}
HMIMO relies on frequency-dependent phase shifts and gains to shape wavefronts. This leads to beam misalignment across bands, which is particularly critical in multi-band NTN links. Frequency-flat hardware and multi-band designs offer promising solutions to improve multi-band performance.
\subsection*{Advanced Channel Models}
HMIMO arrays with closely spaced elements invalidate conventional i.i.d. channel models, requiring approaches that account for mutual coupling. In NTN, long propagation distances and atmospheric effects further complicate modeling. Also, LoS-dominant channels tend to be highly correlated, limiting rank and spatial multiplexing gains. Cooperative multi-transmitter schemes can mitigate low-rank conditions but introduce challenges in synchronization and coordination.
\section*{Conclusion}
We have examined the role of advanced multi-antenna architectures for integrating NTN into next-generation mobile broadband systems. We have outlined the key technical challenges in NTN connectivity and assessed conventional MIMO-based NTN coverage models. Our assessment shows that their dependence on large numbers of antenna elements and RF chains leads to significant power consumption and hardware complexity, two critical limitations to power- and space-constrained NTN platforms. We advocate for HMIMO as a promising alternative, highlighting its ability to enable more power-efficient beamforming and simpler processing components. The case study demonstrates that integrating HMIMO into LEO-based systems is feasible and can support multi-user connectivity with reasonable performance. LEO constellations are expected to serve as the primary platform enabling NTN integration in IMT-2030 systems. Overall, while HMIMO does not fully resolve all NTN challenges, it provides a path toward practical antenna solutions that enable NTN–terrestrial integration in future mobile broadband systems.
\bibliographystyle{IEEEtran}
\bibliography{references}

@misc{ITU2023,
      title={Framework and Overall Objectives of the Future Development of \uppercase{IMT} for 2030 and Beyond (\uppercase{ITU-R M.2160-0} Report)}, 
      author={ITU},
      year={2023},
}

@ARTICLE{Li2023,
  author={Li, Ke-Xin and Gao, Xiqi and Xia, Xiang-Gen},
  journal={IEEE Transactions on Wireless Communications}, 
  title={Channel Estimation for LEO Satellite Massive MIMO OFDM Communications}, 
  year={2023},
  volume={22},
  number={11},
  pages={7537-7550},
  doi={10.1109/TWC.2023.3252895}}

@ARTICLE{Kim2025,
  author={Kim, Donghyeon and Jung, Haejoon and Lee, In-Ho and Niyato, Dusit},
  journal={IEEE Internet of Things Journal}, 
  title={Multibeam Management and Resource Allocation for LEO Satellite-Assisted IoT Networks}, 
  year={2025},
  volume={12},
  number={12},
  pages={19443-19458},
  doi={10.1109/JIOT.2025.3542238}}

@ARTICLE{An2026,
  author={An, Hao and Guan, Ke and He, Danping and Liu, Ting and Mathiopoulos, P. Takis and Duo, Hao and Taheri, Sohail and Ji, Yilin},
  journal={IEEE Network}, 
  title={Channel Modeling for Space-Aerial-Terrestrial Integrated Networks (SATIN)}, 
  year={2026},
  volume={40},
  number={1},
  pages={79-87},
  doi={10.1109/MNET.2025.3579875}}

@ARTICLE{Ran2025,
  author={Ran, Yongyi and Ding, Yajie and Chen, Shuangwu and Lei, Jizhao and Luo, Jiangtao},
  journal={IEEE Transactions on Vehicular Technology}, 
  title={Fully-Distributed Dynamic Packet Routing for LEO Satellite Networks: A GNN-Enhanced Multi-Agent Reinforcement Learning Approach}, 
  year={2025},
  volume={74},
  number={3},
  pages={5229-5234},
  doi={10.1109/TVT.2024.3499933}}

@INPROCEEDINGS{Sun2024,
  author={Sun, Hongbing and Zhang, Qiang and Sun, Lei and Yu, Daqun and Li, Jianxin and Wang, Hua},
  booktitle={2024 Photonics \& Electromagnetics Research Symposium (PIERS)}, 
  title={The Multi-beam Technology in LEO Satellite Communications}, 
  year={2024},
  volume={},
  number={},
  pages={1-5},
  doi={10.1109/PIERS62282.2024.10618424}}

@ARTICLE{Li2022,
  author={Li, Ke-Xin and You, Li and Wang, Jiaheng and Gao, Xiqi and Tsinos, Christos G. and Chatzinotas, Symeon and Ottersten, Björn},
  journal={IEEE Transactions on Communications}, 
  title={Downlink Transmit Design for Massive MIMO LEO Satellite Communications}, 
  year={2022},
  volume={70},
  number={2},
  pages={1014-1028},
  doi={10.1109/TCOMM.2021.3131573}}

@ARTICLE{Abdelsadek2022,
  author={Abdelsadek, Mohammed Y. and Kurt, Gunes Karabulut and Yanikomeroglu, Halim},
  journal={IEEE Open Journal of the Communications Society}, 
  title={Distributed Massive MIMO for LEO Satellite Networks}, 
  year={2022},
  volume={3},
  number={},
  pages={2162-2177},
  doi={10.1109/OJCOMS.2022.3219419}}

@ARTICLE{Li2025,
  author={Li, Qingchao and El-Hajjar, Mohammed and Cao, Kaijun and Xu, Chao and Haas, Harald and Hanzo, Lajos},
  journal={IEEE Transactions on Wireless Communications}, 
  title={Holographic Metasurface-Based Beamforming for Multi-Altitude LEO Satellite Networks}, 
  year={2025},
  volume={24},
  number={4},
  pages={3103-3116},
  doi={10.1109/TWC.2025.3527962}}

@ARTICLE{Zhang2025,
  author={Zhang, Xin and Qin, Xiaohan and Zhang, Zitian and Cai, Lin X. and Zhou, Haibo and Zhuang, Weihua},
  journal={IEEE Internet of Things Journal}, 
  title={RIS-Aided MIMO Downlink Transmission for Ultradense LEO Satellite-Terrestrial Networks}, 
  year={2025},
  volume={12},
  number={11},
  pages={15304-15318},
  doi={10.1109/JIOT.2025.3532342}}

@ARTICLE{Huang2020,
  author={Huang, Chongwen and Hu, Sha and Alexandropoulos, George C. and Zappone, Alessio and Yuen, Chau and Zhang, Rui and Renzo, Marco Di and Debbah, Merouane},
  journal={IEEE Wireless Communications}, 
  title={Holographic MIMO Surfaces for 6G Wireless Networks: Opportunities, Challenges, and Trends}, 
  year={2020},
  volume={27},
  number={5},
  pages={118-125},
  doi={10.1109/MWC.001.1900534}}

@ARTICLE{11195786,
  author={Hossain, Ekram and Vera-Rivera, Angelo},
  journal={IEEE Transactions on Technology and Society}, 
  title={6G Cellular Networks: Mapping the Landscape for the IMT-2030 Framework}, 
  year={2025},
  volume={},
  number={},
  pages={1-16},
  doi={10.1109/TTS.2025.3611364}}

@ARTICLE{Gong2024,
  author={Gong, Tierui and Gavriilidis, Panagiotis and Ji, Ran and Huang, Chongwen and Alexandropoulos, George C. and Wei, Li and Zhang, Zhaoyang and Debbah, Mérouane and Poor, H. Vincent and Yuen, Chau},
  journal={IEEE Communications Surveys \& Tutorials}, 
  title={Holographic MIMO Communications: Theoretical Foundations, Enabling Technologies, and Future Directions}, 
  year={2024},
  volume={26},
  number={1},
  pages={196-257},
  doi={10.1109/COMST.2023.3309529}}

@ARTICLE{An2023,
  author={An, Jiancheng and Xu, Chao and Ng, Derrick Wing Kwan and Alexandropoulos, George C. and Huang, Chongwen and Yuen, Chau and Hanzo, Lajos},
  journal={IEEE Journal on Selected Areas in Communications}, 
  title={Stacked Intelligent Metasurfaces for Efficient Holographic MIMO Communications in 6G}, 
  year={2023},
  volume={41},
  number={8},
  pages={2380-2396},
  doi={10.1109/JSAC.2023.3288261}}

@ARTICLE{8910627,
  author={Wu, Qingqing and Zhang, Rui},
  journal={IEEE Communications Magazine}, 
  title={Towards Smart and Reconfigurable Environment: Intelligent Reflecting Surface Aided Wireless Network}, 
  year={2020},
  volume={58},
  number={1},
  pages={106-112},
  doi={10.1109/MCOM.001.1900107}}

\begingroup
\let\description\LaTeXdescription
\let\enddescription\endLaTeXdescription
\endgroup
%




\end{document}